\documentstyle[aps,epsf]{revtex}

\begin{document}
\title{Phase space density and chiral symmetry restoration in
relativistic heavy ion collisions}
\author{Scott Pratt and Kevin Haglin\footnote{Current address: Department of
Physics, Astronomy and Engineering Science, St. Cloud State University,
St. Cloud, MN  56301}}
\address{Department of Physics and Astronomy and \\
National Superconducting Cyclotron Laboratory, \\
Michigan State University, East Lansing, MI 48824~~USA}
\date{\today}
\maketitle

\begin{abstract}
The effect of altered hadron masses is studied for its effect with regard to
final-state hadronic observables. It is shown that the final phase space
densities of pions and kaons, which can be inferred experimentally, are
sensitive to in-medium properties of the excited matter at earlier stages of
the collision, but that the sensitivity is significantly moderated by
interactions that change the effective numbers of pions and kaons during the
latter part of the collision.
\end{abstract}

\pacs{25.75.Gz,25.75.-q,25.75.Dw}


Relativistic heavy ion collisions at the CERN SPS (160$A$ GeV Pb + Pb) or the
Brookhaven AGS (11$A$ GeV Au + Au) produce a mesoscopic region where initial
energy densities are in the neighborhood of a few
GeV/fm$^3$\cite{wa98_qm96,na49_qm95}, several times the energy density of a
proton. As the system expands and cools normal hadronic degrees of freedom
become justified, but the system might still have novel properties when energy
densities are in the region of 500 MeV/fm$^3$. Most notably, as a consequence
of restoring chiral symmetry, the masses of heavier hadrons might fall as much
as 50\%\cite{brownrho}. Evidence of falling masses has appeared in dilepton
measurements of the $\rho$ peak in Pb + Pb collisions at $E/A=160$ GeV,
performed at the SPS. The observed lack of strength of dilepton pairs in the
region of the $\rho$ meson mass, combined with the observed additional strength
for invariant masses near 400 MeV, suggest either that the $\rho$ meson was
altered\cite{likobrown}, dissolved\cite{siemenschin}, or broadened beyond
recognition by collisions\cite{wambaugh}.

Since dileptons largely pass through the collision volume unscathed, they
provide a transparent probe for investigating hadronic properties during the
most interesting stages of the collision, but only those hadrons with the
quantum numbers of the photon. Final state measurements of hadrons -- pions,
kaons, protons and hyperons, provide a rich chemistry as nearly $10^3$ hadrons
are commonly produced in a single event. However, hadrons interact several
times after the system has expanded beyond the energy densities of greatest
interest when temperatures are near or above 150 MeV, and before the breakup
density is reached, when temperatures are approximately 110 MeV. In the high
density hadronic state, equilibrium chemical abundances can easily change by
factors of two if masses are adjusted by several hundred MeV as predicted by
some chiral models. In this letter we investigate whether such mass changes,
and the corresponding changes in hadronic chemical abundances, survive and
produce a signal in the measured hadrons. By using phenomenologically motivated
chemical rates, we model the chemical development of a kinetically equilibrated
gas that expands under a time scale relevant for relativistic heavy ion
collisions. We find that falling hadron masses, on the order of 50\%, result in
increased final phase space densities for both pions and kaons which can be
inferred by measuring spectra and two-particle correlations. We show that if
reactions that preserve the effective pion and kaon numbers are ignored, a
strong signal survives the final expansion, and that if such reactions are
included, the manifestations of the original novel chemistry survive, but are
strongly moderated. We also find that hadronic observables are sensitive to the
issue of whether both baryon and vector mesons scale, or whether just the
baryon masses\cite{pisarski,koch}.

In Pb + Pb collisions at the SPS or in Au +Au collisions at the AGS, $\approx$
7 fm/c after initial contact, matter has expanded and cooled to the point where
a model based on binary interactions of hadrons is justified.  For the
following thermal calculations, we take the onset of binary modeling to occur
at a temperature of 160 MeV. We assume that the system is both kinetically and
chemically equilibrated at this point and is characterized by four numbers: a
baryon chemical potential $\mu_b$, a strangeness chemical potential $\mu_s$,
and a temperature $T$. The strangeness chemical potential is chosen such that
the net strangeness is zero. It would be zero if the baryon density were
zero. The baryon chemical potential is chosen to match the effective baryon to
pion ratios, unity for the AGS example and 1/5 for the SPS example.  The
assumption of equilibration when $T=160$ MeV is an ansatz, justified only by
the fact that the hadrons significantly overlap at higher temperature,
suggesting very rapid equilibration of the strongly interacting system. At
later times and lower temperatures, kinetic equilibrium is approximately
maintained\cite{bnc}, but some aspects of chemical equilibrium are lost. When
the system expands and cools to approximately $T=110$ MeV, kinetic equilibrium
is lost as well and the system dissolves.

As the system expands, the effective numbers of kaons and pions must adjust to
the rapidly changing environment. The effective number of pions is defined as
the number of pions plus the sum of other hadrons weighted by their effective
pionic content. For instance, a $\rho$ meson counts as two pions since it
decays principally into two pions, while the effective pionic content of a
$\Delta$ baryon is one. Reactions such as $\rho\leftrightarrow\pi\pi$ are rapid
and can keep the number of $\rho$ mesons in equilibrium but do not change the
overall effective pion number\cite{song,gavin}. Reactions that change the
overall pion number, such as $\pi\pi\leftrightarrow\rho\rho$, are not
sufficiently fast and allow the system to lose chemical equilibrium. One can
characterize the unequilibrated state by an effective chemical potential
$\mu_\pi$ that corresponds to the ``conserved'' pion number. One can similarly
assign an effective chemical potential to the effective kaon number. Although
the net strangeness is zero, the net number of strange quarks must adjust
during the expansion. Reactions that conserve the net number of strange quarks
occur rapidly, e.g. $K^-p\leftrightarrow\Lambda\pi$, but reactions that change
the number of strange quarks, e.g. $\pi\pi\leftrightarrow KK$, are slow as they
require a strange and an anti-strange hadron to either interact or be be
produced jointly. We again describe this lack of chemical equilibrium via an
effective chemical potential $\mu_K$\cite{rafelski} that corresponds to the
effective kaon number, where a kaon or $\Lambda$ baryon counts as one kaon as
they each have one strange quark. Either an $\Omega$ baryon (with quark content
$sss$) or an $\bar{\Omega}$ would count as three kaons. Both $\mu_K$ and
$\mu_\pi$ become zero in chemically equilibrated systems.

When a system is perturbed from chemical equilibrium, $\mu_K$ and $\mu_\pi$
approach equilibrium exponentially with characteristic times $\tau_K$ and
$\tau_\pi$. These times have been studied by Song and Koch for a simple meson
gas\cite{song}. Our analysis includes a greater variety of hadrons. We include
the spin 1/2 baryon octet, the baryon spin 3/2 dectet, the pseudoscalar meson
nonet and the vector meson nonet, but not those resonances which correspond to
orbital excitations of mesons in the framework of the constituent quark
model. Given the cross sections for creating pions in individual reactions, we
estimate the time reversed rate through time-reversal arguments.
\begin{eqnarray}
\frac{dN_{\pi}}{d^3xdt}&=&\sum_i \Delta N_{i,\pi}
R_i (1-\exp{-\Delta N_{i,\pi}\mu_pi}),\\
\nonumber
R_i&=&\frac{(2J_a+1)(2J_b+1)}{(2\pi)^6}\int \frac{E_{cm} d^3P}{E}
e^{-[E-(N_{a,\pi}+N_{b,\pi})\mu_{\pi})]/T}
\int d^3q \sigma(E_{cm})
v_{rel,cm}
\end{eqnarray}
where the sum is over all reactions ($ab\rightarrow X$) that increase the
effective pion number by $\Delta N_{i,\pi}$. The effective rate per volume,
$R_i$, for a specific reaction can be found by convoluting the phase space
densities of $a$ and $b$, and folding in the cross section.The energy of the
pair is $E=\sqrt{P^2+E_{cm}^2}$, where $E_{cm}$ is the energy of the pair as
measured in the $ab$ rest frame and $q$ and $v_{rel,vm}$ are the relative
momentum and relative velocity of $a$ and $b$ as measured in that frame. The
baryon and strangeness chemical potentials have been omitted for brevity. The
integral over $P$ can be performed analytically, and given an expression for
the cross section, the integral over $q$ can be performed numerically.

The effective pion density,
\begin{equation}
n_{\pi}=\sum_a N_{a,\pi}(2J_a+1)\int \frac{d^3p}{(2\pi)^3}
e^{-(E-\mu_{\pi}N_{a,\pi})},
\end{equation}
can also be found analytically given the effective pionic content, $N_{a,\pi}$
of the species $a$.  One can expand the above expressions for small $\mu_{\pi}$
to find the rate at which $\mu_{\pi}$ returns to equilibrium,
\begin{eqnarray}
\frac{d\mu_\pi}{dt}&=&-\frac{1}{\tau}\mu_\pi\\
\nonumber
\frac{1}{\tau}&=&\frac{\sum_i \Delta N_{i,\pi}^2
R_i(\mu_\pi=0)}{n_{\pi}(\mu_\pi=0)}
\end{eqnarray}

Unfortunately, experimental cross sections are not available for the majority
of the combinations $ab$ of the 26 mass states used in this analysis, and since
the relevant energies for pion production are several hundred MeV, perturbation
theory is not particularly reliable. We have therefore instituted simplified
expressions for the cross sections.
\begin{equation}
\sigma(E_{cm})=\sigma_0 \theta(E_{cm}-E_{th}-N_{\pi,X} 350~ {\rm MeV}),
\end{equation}
where $N_{\pi,X}$ is the number of pions in the final state, and the threshold
$E_{th}$ is the minimum energy of a state with the quantum numbers of the
initial state. This behavior is motivated by the observed behavior of the
inelastic $pp$ and $p\pi$ cross sections\cite{partdatbook} with $\sigma_0$
being 25 mb if both incoming particles are baryons, 20 mb if one is a baryon
and 15 mb if neither is a baryon. When the production of more pions becomes
available, all the cross section is then devoted to that number of pions. For
the case where no baryons are present in the initial state, only even numbers
of pions are allowed in the final state as required by conservation of
g-parity. For the production of strangeness, only pairs of strange-anti strange
are allowed and the steps are in units of 1.2 GeV if no baryons are present in
the initial state and 800 MeV if baryons are present.
\begin{equation}
\sigma(E_{cm})=\sigma_0 \theta\left(E_{cm}-E_{th}-N_{K,X}(800{\rm ~or~}1200~{\rm MeV})\right),
\end{equation}
The reduced stepsize of 800 MeV is to account for $\Lambda$ production which
only requires a few hundred MeV energy. The cross section $\sigma_0$ for
strangeness production is taken as one fourth that of the pion production cross
section. These values were motivated by measurements of $\Lambda$
production\cite{sibirtsev,lambdaprod}, with the value of $\sigma_0$ being
multiplied by 2.5 to account for the production of other hyperons.  Although
the prescription for determining cross sections is phenomenologically
motivated, given the large number of experimentally unknown information, the
cross sections must not be taken seriously beyond the 50\% level, especially
for the case of strangeness producing rates.

Characteristic chemical equilibration times are displayed in Figure
1. Characteristic expansion times are $\approx 5$ fm/c when $T=160$ MeV and
$\approx 20$ fm/c when the system is near breakup at $T\approx 110$ MeV. From
viewing Figure 1 one expects pionic chemical equilibrium to
be lost when the temperature approaches 150 MeV while strangeness equilibration
can be hardly justified even at $T=160$ MeV. We should point out that if masses
fall due to restoration of chiral symmetry, hadron densities rise accordingly
and characteristic times are considerably shorter. The role of the baryons in
maintaining chemical equilibrium is crucial. Even when baryons comprise only a
fifth of the produced hadrons, reactions involving baryons contribute the
majority of the rate. This comes from the fact that reactions involving baryons
do not need to conserve g-parity and pions need not be produced pairwise.

We now proceed to calculate generated chemical potentials for systems which
cool from an equilibrated state at $T=160$ MeV to an unequilibrated state at
$T=110$ MeV with chemical potentials $\mu_{\pi}$ and $\mu_K$. If
number-changing rates are neglected the four final chemical potentials,
$\mu_{\pi}$, $\mu_K$, $\mu_s$ and $\mu_b$ can be found by the four constraints:
(1) The net strangeness is zero. (2) The baryon to pion ratio is fixed. (3) The
entropy per pion is fixed. (4) The net number of strange quarks per pion is
fixed. If rates that change the pion number and strange quark number are
included, the latter three constraints must be modified by integrating the time
development of the system.
\begin{eqnarray}
\frac{d}{dt} \frac{n_b}{n_{\pi}} &=&
-\frac{n_b}{n_{\pi}^2}\frac{d}{dt} n_{\pi}\\
\frac{d}{dt} \frac{n_S}{n_\pi} &=&
-\frac{n_S}{n_{\pi}^2}\frac{d}{dt} n_{\pi}
-\frac{\mu_{\pi}}{n_{\pi}} \frac{d}{dt} n_{\pi}
-\frac{\mu_K}{n_{\pi}} \frac{d}{dt} n_{K} \\
\frac{d}{dt} \frac{n_K}{n_{\pi}} &=&
-\frac{n_K}{n_{\pi}^2}\frac{d}{dt} n_{\pi}
+\frac{1}{n_{\pi}}\frac{d}{dt} n_K,
\end{eqnarray}
where $n_S$ is the entropy density. Thus if one knows the temperature as a
function of time, one can integrate these equations forward in time, using the
four aforementioned quantities to determine the four unknown chemical
potentials at any time. For our purposes we assumed a simplified behavior of
the temperature, $T$, as a function of time, $dT/dt$ = -6.5 MeV/(fm/c), which
was motivated by the behavior observed in cascade simulations\cite{bnc}.

The resulting evolutions of $\mu_\pi$ and $\mu_K$ as a function of temperature
are illustrated in Figure 2. The baryon to pion ratio was chosen to be 0.2,
which is relevant for Pb + Pb collisions at the SPS. Inclusion of the rates is
clearly important in determining the final chemistry.

The chemical evolutions were also calculated with the assumption that the
hadronic masses varied as a function of the temperature. The hadrons, aside
from the pseudoscalar mesons which are Goldstone bosons, were assumed to scale
linearly with the temperature from their vacuum mass when $T=110$ MeV to a
fraction, $m/m_0$, at $T=160$ MeV. The lower panels of Figure 3 show the value
of $\mu_{\pi}$ at breakup for four cases. The final baryon to pion ratio was
chosen to be 0.2 or 1.0, roughly appropriate for SPS or AGS conditions
respectively. If no number-changing rates are included, the generated chemical
potential is large, approaching 100 MeV for the SPS case and 150 MeV for the
AGS case, when the mass reduction factor falls below 0.5. However, inclusion of
the rates reduces the resulting chemical potential to near 50 MeV. Since the
falling of the $\rho$ mass is controversial, the calculations were repeated
with the assumption that only the baryon masses scaled (right panels of Figure
3), while the vector meson masses remained fixed. In this case the resulting
chemical potentials were far lower for the SPS case, and in fact chemical
potentials were smaller for increasingly small mass reduction factors. This
owes itself to the fact that the entropy per pion due to the presence of
baryons is rather high, compared to the entropy per pion in the mesonic sector.

Calculations of $\mu_K$ as a function of the mass reduction factor are
displayed in the upper panels of Figure 3. The resulting $\mu_K$ is much larger
than $\mu_\pi$, surpassing 100 MeV, even for the case where rates were
included. Since the phase space density is proportional to $e^{\mu/T}$, the
strangeness phase space density is nearly doubled compared to the $\mu_K=0$
case if masses don't scale, and more than doubled when mass-scaling occurs. The
number-changing rates that most strongly affected $\mu_K$ were not those that
changed the effective kaon number but those that affected the net pion
number. The kaon-number-changing rates are sufficiently small that they had a
relatively small effect toward the final outcome.

Chemical potentials can be inferred from hadronic measurements via correlation
measurements\cite{bertsch}. Combining two-particle correlations measurements,
which are sensitive to the breakup volume, and spectra one can infer phase
space densities, which should be $\approx e^{\mu/T}$, at low momentum. Combined
with careful modeling of the breakup stage of the reaction, one can thus infer
chemical potentials to the accuracy of $\pm 30$ MeV.

Perhaps the observable that is most directly sensitive to $\mu_K$ is the
abundance of anti-hyperons. For instance, $\bar{\Omega}$ consists of three
anti-quarks, and would feel an enhancement of $\exp{(3\mu_K/T)}$. Figure 4
displays the ratio $\bar{\Lambda}/\bar{p}$ as a function of the mass reduction
factor. The baryon-to-pion ratio was set to 0.2 or 1.0 to be relevant for the
SPS and AGS experiments respectively. Calculations where number changing rates
are included or not included are both shown, and it is evident that
number-changing rates have a profound influence on the population of
anti-hyperons. When rates are included, there is little sensitivity to the mass
reduction factor, whereas when number changing rates are ignored, calculations
yield very large ratios. Preliminary measurements of the
$\bar{\Lambda}/\bar{p}$ by E864\cite{e864lambdabar} have yielded values above
unity, approximately double or triple the calculations presented here. If one
argued that the cross sections for number changing rates used here,
particularly those that changed the effective number of strange quarks, were
overestimated by a factor of two, one could then explain the experimental
findings. Before one can make firm conclusions with regard to this measurement,
a better understanding of number-changing cross sections, especially those
involving the annihilation of strange hyperons, must be reached.

Several conclusions can be made from this investigation. First, the chemical
properties of the hottest stages of relativistic heavy ion collisions do indeed
manifest themselves in the final hadronic measurements. In particular, chiral
restoration can affect the final phase space densities of both the pions and
kaons. However, chemical processes, which drive the system towards equilibrium,
significantly mitigate such signals. Altering the chiral properties of the
hottest stages results in a difference of the final chemical potentials of the
order of 50 MeV or less, while the ability of an experiment to determine the
chemical potentials, is probably in the 30 MeV range at best. Thus, measuring
hadrons can be expected to provide information or evidence regarding the chiral
transition, but would probably not supply a ``smoking gun'' signal, unless the
chemical rates employed in this paper are overestimated.

It should be stressed that this analysis has relevance beyond the issue of
dropping hadron masses. Effective chemical potentials and phase space densities
are of crucial importance in understanding the entropy of the reaction, which
can signal the presence of the deconfinement transition. Secondly, the
conditions of the final state provide boundary conditions necessary for
modeling the earlier stages of the collision. For instance, dilepton production
from $\pi\pi$ annihilation would scale as $\exp{(2\mu_{\pi}/T)}$. Finally, the
sensitivity of the observables to the inclusion of pion-number-changing and
kaon-number-changing chemical processes demonstrates the importance of
including such reactions, along with the corresponding time-reversed processes,
in any modeling of the final state expansion.

\acknowledgements{The authors thank Yang Pang for pointing out the importance
of baryons in hadronic chemistry. This work was supported by the National
Science Foundation under grant PHY-9605207. }


\newpage

\begin{figure}
\label{equiltimes_fig}
\centerline{\epsfxsize=7.0cm \epsffile{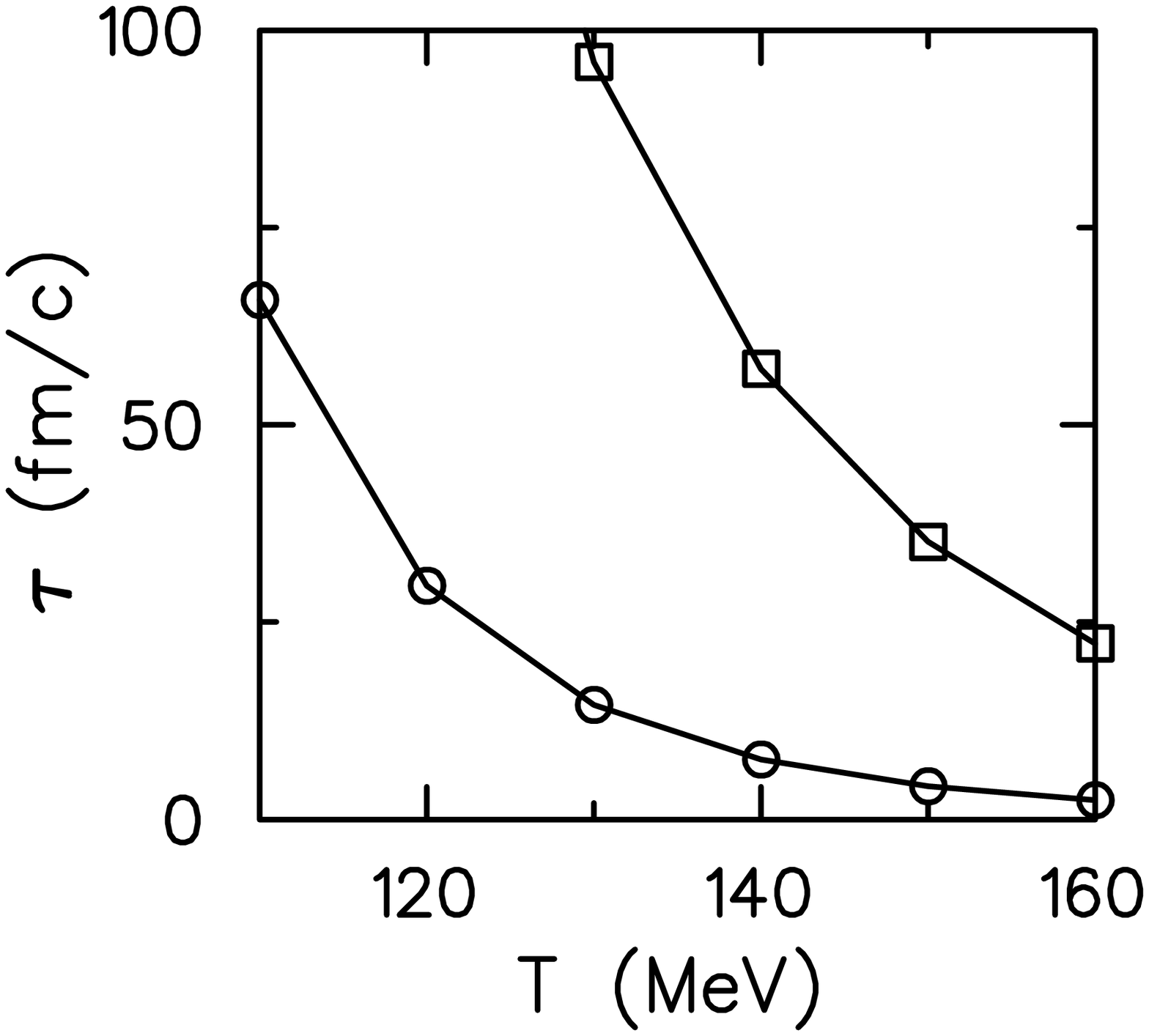}}
\caption{Characteristic times for chemical potentials returning to zero are
shown for $\mu_{\pi}$ (circles) and $\mu_K$ (squares). The decay time
corresponding to $\mu_{\pi}$ becomes larger than characteristic expansion times
when $T \approx 150$ MeV, while equilibration times for $\mu_K$ are longer than
characteristic expansion times even for $T=160$ MeV.}
\end{figure}

\begin{figure}
\label{muvstemp_fig}
\centerline{\epsfxsize=7.0cm \epsffile{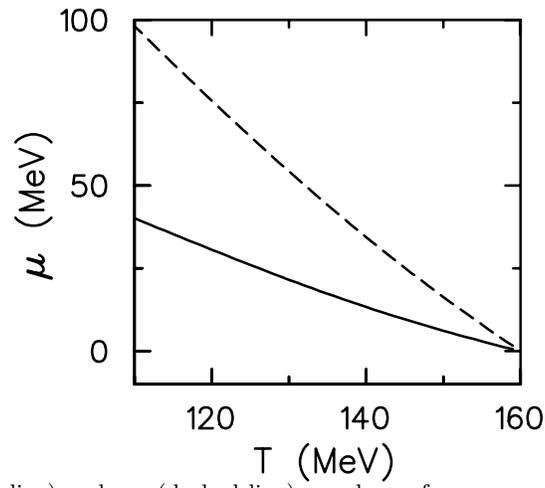}}
\caption{The evolution of $\mu_{\pi}$ (solid line) and $\mu_K$ (dashed line)
are shown for an expanding hadron gas that began at a 160 MeV temperature. This
calculation has incorporated pion and kaon number changing rates.}
\end{figure}

\begin{figure}
\label{muvsscale_fig}
\centerline{\epsfxsize=9.0cm \epsffile{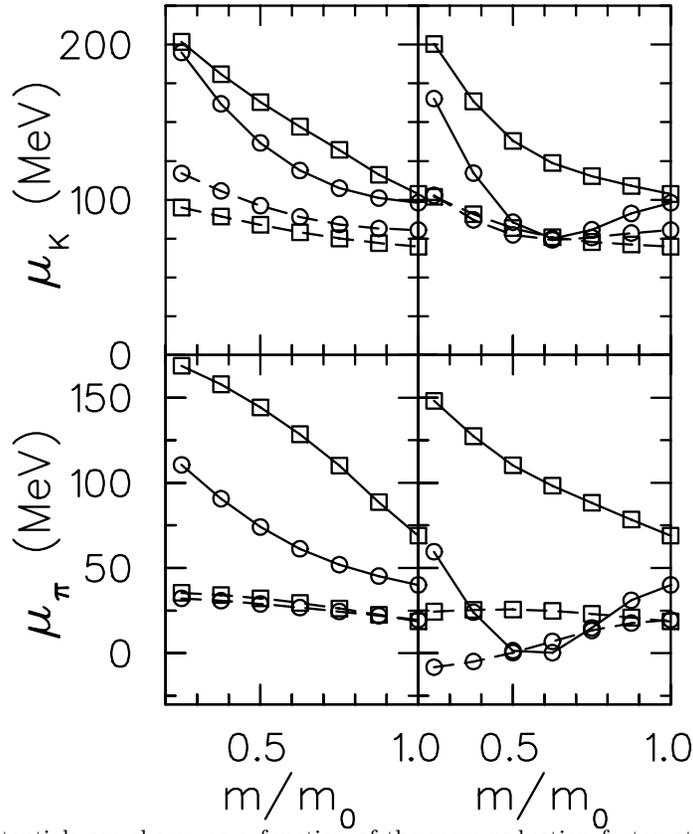}}
\caption{ Final chemical potentials are shown as a function of the mass
reduction factor at the initial temperature of 160 MeV. Calculations are shown
for the cases where number-changing processes are neglected (solid lines) or
included (dashed lines), and for when vector mesons scale with the baryon
masses (left panel) or remain fixed (right panel). Calculations are performed
for two choices of the baryon-to-pion ratio, 1.0 which is relevant for AGS
measurements (squares) and 0.2 which is relevant for measurements at SPS
(circles). One sees that number-changing rates significantly damp the effect of
altering hadron masses, and if vector meson masses are unchanged that the
chemistry is less affected.}
\end{figure}

\begin{figure}
\label{lambdabar_fig}
\centerline{\epsfxsize=7.0cm \epsffile{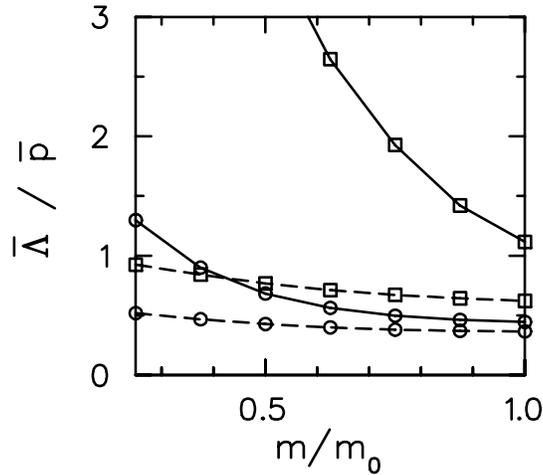}}
\caption{The $\bar{\Lambda}/\bar{p}$ ratio is plotted as a function of the
mass reduction factor. Calculations are shown for the cases where
number-changing processes are neglected (solid lines) or included (dashed
lines). Calculations are performed for the two choices of the baryon-to-pion
ration, 1.0 which is relevant for AGS measurements (squares) and 0.2 which is
relevant for measurements at CERN (circles).  Calculations are sensitive
to both the inclusion of number-changing rates and alterations of hadronic
masses.}
\end{figure}

\end{document}